\documentclass[preprint,prl,superscriptaddress,showpacs,preprintnumbers,
amsmath,amssymb,showpacs,,showkeys]{revtex4}

\usepackage{graphicx}
\usepackage{dcolumn}
\usepackage{bm}
\usepackage{hyperref}
\usepackage{subfigure}
\usepackage{color}
\usepackage{ulem}
\begin{document}

\title{Atomic monolayer deposition on the surface of nanotube mechanical resonators}

\affiliation{ICFO - Institut de Ciencies Fotoniques, Mediterranean Technology 
Park, 08860 Castelldefels, Barcelona, Spain}
\affiliation{Institut Catal{\`a} de Nanotecnologia, Campus de la UAB, 
E-08193 Bellaterra, Spain} 
\affiliation{Departamento de Sistemas F\'{\i}sicos, Qu\'{\i}micos y
 Naturales, Universidad Pablo de Olavide, Carretera de Utrera, km 1,
  E-41013 Sevilla, Spain}
\affiliation{Departament de F\'{\i}sica i Enginyeria Nuclear, 
Universitat Polit\`ecnica de Catalunya, B4-B5 Campus Nord, 08034 Barcelona, Spain}

\author{A. Tavernarakis}
\affiliation{ICFO - Institut de Ciencies Fotoniques, 
Mediterranean Technology Park, 08860 Castelldefels, Barcelona, Spain}

\author{J. Chaste}
\altaffiliation[]{Present address: CNRS, Laboratoire de Photonique
 et de Nanostructures, UPR20, route de Nozay, 91460 Marcoussis, France}
\affiliation{Institut Catal{\`a} de Nanotecnologia, Campus de la UAB,
 E-08193 Bellaterra, Spain} 

\author{A. Eichler}
\altaffiliation{Present address: Department of Physics, ETH Zurich, 
Schafmattstrasse 16, 8093 Zurich, Switzerland.}
\affiliation{ICFO - Institut de Ciencies Fotoniques, Mediterranean 
Technology Park, 08860 Castelldefels, Barcelona, Spain}
\affiliation{Institut Catal{\`a} de Nanotecnologia, Campus de 
la UAB, E-08193 Bellaterra, Spain}

\author{G. Ceballos}
\affiliation{Institut Catal{\`a} de Nanotecnologia, Campus de la UAB,
 E-08193 Bellaterra, Spain} 

\author{M. C. Gordillo}
\affiliation{Departamento de Sistemas F\'{\i}sicos, Qu\'{\i}micos
 y Naturales, Universidad Pablo de Olavide, Carretera de Utrera, km 1,
  E-41013 Sevilla, Spain}

\author{J. Boronat}
\affiliation{Departament de F\'{\i}sica i Enginyeria Nuclear,
 Universitat Polit\`ecnica de Catalunya, B4-B5 Campus Nord, 08034 Barcelona, Spain}

\author{A. Bachtold}
\affiliation{ICFO - Institut de Ciencies Fotoniques, Mediterranean 
Technology Park, 08860 Castelldefels, Barcelona, Spain}
\affiliation{Institut Catal{\`a} de Nanotecnologia, Campus de la UAB,
 E-08193 Bellaterra, Spain} 


\begin{abstract}
We studied monolayers of noble gas atoms (Xe, Kr,
Ar, and Ne) deposited on individual ultra-clean suspended
nanotubes. For this, we recorded the resonance frequency of the
mechanical motion of the nanotube, since it provides a direct
measure of the coverage. The latter is the number of adsorbed
atoms divided by the number of the carbon atoms of the suspended
nanotube. Monolayers formed when the temperature was lowered in a
constant pressure of noble gas atoms. The coverage of Xe
monolayers remained constant at 1/6 over a large temperature
range. This finding reveals that Xe monolayers are solid phases
with a triangular atomic arrangement, and are commensurate with
the underlying carbon nanotube. By comparing our measurements to
theoretical calculations, we identify the phases of Ar and Ne
monolayers as fluids, and we tentatively describe Kr monolayers as
solid phases. These results underscore that mechanical resonators
made from single nanotubes are excellent probes for surface
science.
\end{abstract}
\maketitle

Carbon nanotubes have motivated a considerable research effort for
the study of gas adsorption onto substrates that approach the
one-dimensional limit
\cite{dillon1997storage,teizer19994,ulbricht2002desorption,shi2003gas,
lasjaunias2003evidence,ulbricht2006thermal,Wang:2010hi,Lee:2012ic,Chaste_NatNano12}.
Many studies have been carried out on mats and films of nanotube bundles,
but the interpretation of those measurements is complicated by the
fact that the binding energy of the gas atoms on the substrate is
not homogeneous. That is, the binding energy depends on whether
the atom is located on an individual nanotube, at the junction
between two crossing nanotubes, or along the interstitial channel
formed between two parallel nanotubes. Recently, this homogeneity
problem was solved by studying gas adsorption on individual
nanotubes, a technical feat considering the tiny amount of
adsorbed atoms \cite{Wang:2010hi,Lee:2012ic,Chaste_NatNano12}. For
this, nanotubes were employed both as substrates for adsorption
and as detectors. Namely, the nanotubes were operated as
mechanical resonators, the resonance frequency being exquisitely
sensitive to the number of adsorbed atoms
\cite{Chaste_NatNano12,Chiu:2008eq,lassagne2008ultrasensitive,jensen2008atomic}.

\begin{figure}[t!]
  \centering
  \includegraphics[scale=1]{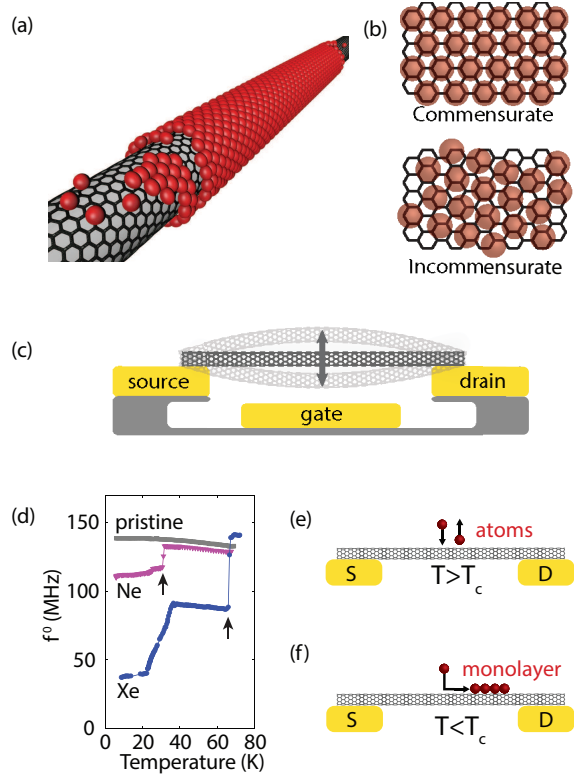}
  \caption{
(a) Growth of an atomic monolayer on a nanotube. (b) Schematics of
monolayers in the solid phase that are commensurate (top) and
incommensurate (bottom) with the carbon substrate.
The adsorbed atoms are represented by red
spheres, whereas the carbon surface is depicted by the honeycomb
lattice. (c) Layout of the nanotube resonator. (d) Resonance
frequency upon lowering temperature while dosing Xe and Ne using a
pinhole micromanipulator. The curve labeled ``pristine"
corresponds to the $T$ dependance of $f^0$ when we do not dose
atoms. The pressure is $3\cdot 10^{-7}$ mbar for the Xe and the Ne
measurements and $3\cdot 10^{-11}$ mbar for the pristine
measurement. (e,f) Schematics showing the balance of atoms
impinging on and departing from the nanotube above and below the
characteristic temperature $T_{\rm c}$.}
  \label{fig:Fig1}
  \end{figure}

Atoms adsorbed on graphitic surfaces can form a rich variety of
different phases, such as vapor, liquid, supercritical fluids, and
solids \cite{bruch1997physical}. The solid phase can be either
commensurate or incommensurate with the graphene surface (Fig.
\ref{fig:Fig1}(a,b)). The commensurate solid phase is robust,
since the crystal formed by the adsorbed atoms is strongly pinned
to the underlying carbon surface. Commensurate monolayers on
graphite feature a well defined ratio between the number of
adsorbed atoms and the number of carbon atoms at the surface. This
ratio, called coverage,  is often 1/6 for noble gas atoms, which
corresponds to a registered $\sqrt{3} \times \sqrt{3}$ lattice
(table 6.1 in Ref. \cite{bruch1997physical}). This particular coverage value arises
because, in this solid phase, noble gas atoms form a
two-dimensional triangular arrangement in which atoms occupy the
center of carbon hexagons, leaving an empty one in the center of
the triangle (see Fig. \ref{fig:Fig1}b). Nanotubes are also
expected to host commensurate solids; however, due to cylindrical
boundary conditions, these solids exist only for some specific
nanotube chiralities $(n,m)$, namely when $(n-m)/3$ is an integer
\cite{Wang:2010hi}. Remarkably, this is also the condition for
nanotubes to be metallic.

Solid and fluid monolayers made of noble gas atoms, such as
Xe, Kr, and Ar, were measured on graphite surfaces only when
the coverage was comparable to or larger than 1/6
(chapter 6 in Ref. \cite{bruch1997physical}). When adsorbed on a nanotube 
surface, the coverage of incommensurate solids and fluids is expected to become larger 
than that measured on graphite due to the curvature of the nanotube 
\cite{Lee:2012ic}. This is because (i)
adsorbed atoms form cylindrical monolayers with a surface larger
than that of the carbon nanotube, and (ii) the two-dimensional
density of noble gas atoms is, to a good approximation, independent
of the curvature of the monolayer. Recently, monolayers of Kr and
Ar were obtained on individual nanotubes by increasing the
pressure of Kr or Ar gas surrounding the nanotube
\cite{Wang:2010hi}. These monolayers were identified as solids,
but these phases were fragile, since the number of atoms in the
monolayer was very sensitive to temperature.

Here, we report on the formation of monolayers of Xe, Kr, Ar, and
Ne on individual ultra-clean nanotubes upon decreasing
temperature.  The pressure was kept constant, typically in the
$10^{-7}$ mbar range. We prepared the nanotube by thoroughly
current annealing it in order to remove contamination from the
surface. The monolayer of Xe was found to be the most robust
phase. Its coverage remained constant at 1/6 over a large
temperature range, indicating the formation of a $\sqrt{3} \times
\sqrt{3}$ commensurate solid. The coverages of the monolayers made
from Kr, Ar, and Ne were less stable against temperature
variations. We compare our experimental findings to theoretical
calculations in order to establish the nature of these phases.

In order to demonstrate commensurate solid phases, we fabricated
resonators based on ultra-clean nanotubes that are metallic. For
this, we used the fabrication process that we described in Ref.
\cite{eichler2011nonlinear}. As shown in Fig.~\ref{fig:Fig1}(c),
the nanotube is contacted by two electrodes and is suspended over
a trench with a gate electrode at the bottom. The nanotube was
grown by chemical vapor deposition in the last step of the
fabrication process in order to reduce contamination
\cite{huttel2009carbon} (Supplemental Material, Sec. I). The
measurement of the electrical conductance as a function of the
voltage applied to the gate electrode allowed us to select
nanotubes that are metallic with a small energy gap (Supplemental
Material, Sec. III).

The mechanical motion was driven and detected using the
frequency-modulation mixing technique \cite{Barois_Small10}
(Supplemental Material, Sec. III). We carried out the experiment
in a home-built ultra-high vacuum cryostat that reaches a base
pressure of $ \sim3\cdot10^{-11}$ mbar. The nanotube was cleaned
by current annealing. Noble gas atoms were dosed from a
room-temperature supply with a pinhole microdoser. We studied 3
nanotubes yielding similar results. We discuss in the following
the data for one device. Data for a second device are shown in
Supplementary Material, Sec. VIII.

Monolayers of noble gas atoms formed on the nanotube when the
temperature ($T$) was lowered while keeping a constant pressure of
noble gas in the cryostat chamber. The formation was monitored by
measuring the resonance frequency $f^0$ (that is,
by continuously recording the response of the nanotube resonator 
to the driving frequency). Figure \ref{fig:Fig1}(d)
shows prominent jumps of $f^0$ to lower frequencies upon lowering
T (see arrows), indicating the sudden adsorption of a large
quantity of atoms onto the nanotube. For comparison, when we did
not dose atoms, the temperature dependence of $f^0$ is weak and
monotonic (grey curve labeled ``pristine" in Fig.
\ref{fig:Fig1}(d)). This weak dependence is attributed to the thermal
expansion of the electrodes which modifies the spring constant of
the nanotube resonator \cite{Chaste:2011bj}. The coverage at $T$
is extracted using
 \begin{equation}\label{eq:coverage}
 \varphi (T)=\frac{N_{\rm ads}(T)}{N_{\rm
C}}=\frac{m_{\rm C}}{m_{\rm ads}}\Bigg[A\cdot\Bigg(\frac{f_{\rm
prist}^0(T)}{f_{\rm ads}^0(T)}\Bigg)^2-1\Bigg],
 \end{equation}
where $N_{\rm C}$ is the number of C atoms of the suspended nanotube,
$N_{\rm ads}$ is the number of adsorbed atoms on the nanotube, and
$m_{\rm C}$ and $m_{\rm ads}$ are the atomic masses of carbon and
adsorbed atoms, respectively. Here, $f_{\rm ads}^0$ is the
resonance frequency when dosing atoms for adsorption, and $f_{\rm
prist}^0$ is the resonance frequency when not dosing atoms and
keeping the nanotube pristine. The constant $A$ is introduced to
account for variations in the spring constant between the
measurement of $f_{\rm ads}^0(T)$ and that of $f_{\rm
prist}^0(T)$; indeed, the spring constant can be different, if for
instance the gate voltage applied in the measurement of $f_{\rm
ads}^0(T)$ differs from that of $f_{\rm prist}^0(T)$ (Supplemental
Material, Sec. IV). The constant $A$ is fixed so that $\varphi =0$
at high $T$. In Eq. \ref{eq:coverage}, we assume that the spring
tension is insensitive to the tension induced by the interaction
between noble gas atoms, which is two orders of magnitude weaker
than that of covalent C-C bonds \cite{Wang:2010hi}.

\begin{figure}[t!]
   \centering
  \includegraphics[scale=1]{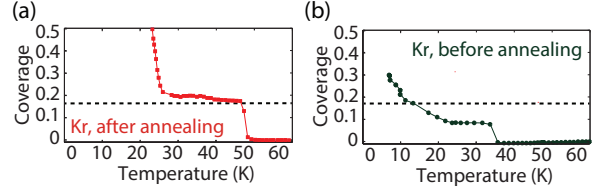}
   \caption{
  (a) Coverage upon dosing Kr atoms
  while lowering the temperature $T$ with a ramping rate $0.016\;\rm K/ s$.
 (b) Same measurement recorded before having current annealed the nanotube 
 with a $T$ ramping rate $0.033\;\rm K/ s$. The pressure is $3\cdot 10^{-7}$ 
 mbar for both measurements.}
   \label{fig:anneal}
   \end{figure}

Figure \ref{fig:anneal}(a) shows the temperature dependence of the
coverage while dosing Kr atoms. Above a characteristic temperature
$T_{\rm c}\simeq 48$~K, the coverage remains at zero. On lowering
temperature, the coverage jumps to  $\varphi\simeq1/6$ and remains
close to this value until $T\simeq26$~K. This behavior can be
accounted for by the balance of atoms impinging on and departing
from the nanotube. For $T>T_{\rm c}$, an impinging atom departs
very rapidly from the nanotube, so that the number of adsorbed
atoms remains close to zero (Fig. \ref{fig:Fig1}(e)). For
$T<T_{\rm c}$, it is energetically favourable for the atoms
to stay on the nanotube (Fig. \ref{fig:Fig1}(f)) -- the atoms
forming a layer with $\varphi\simeq1/6$. This layer is likely 
a monolayer, because the coverage $\varphi\simeq1/6$ of Kr 
on graphite corresponds to a monolayer (chapter 6 in 
Ref. \cite{bruch1997physical}). Upon further lowering 
temperature so that $T<<T_{\rm c}$, the coverage gets larger
than $\varphi=1/6$, indicating that Kr atoms start 
to form the second layer. The coverage grows in a monotonic
way without any additional 
plateaus even when the coverage gets larger than one 
(Supplementary Material, Sec. VII). The absence of additional 
plateaus above $\varphi=1/6$ further supports the interpretation
of the coverage $\varphi\simeq1/6$ as being related to the monolayer.

Key to this work is annealing the nanotube by passing a large
current through it. After the measurements shown in Fig. 2(a), we
exposed the nanotube to ambient air. We then baked the cryostat
and the nanotube at 110~$^\circ$C under vacuum for two days to
reach a base pressure of $ \sim 3\cdot10^{-11}$ mbar. We again
measured the coverage upon lowering T while dosing Kr atoms.  Fig.
\ref{fig:anneal}(b) shows that $T_c$ is much lower than before,
and the coverage at $T\lesssim T_{\rm c}$ is significantly lower
than 1/6. We had to anneal the nanotube with a current of
$\sim10\;\mu\rm A$ in order to recover the same measurement as in
Fig. \ref{fig:anneal}(a). These measurements suggest that the
growth of monolayers is extremely sensitive to contamination,
since a simple exposure to air prevents the formation of
homogeneous monolayers. Another advantage of current annealing is
that it brings the nanotube back to its pristine state after the
adsorption of noble gas atoms on its surface.

\begin{figure}[t!]
      \centering
    \includegraphics[scale=1]{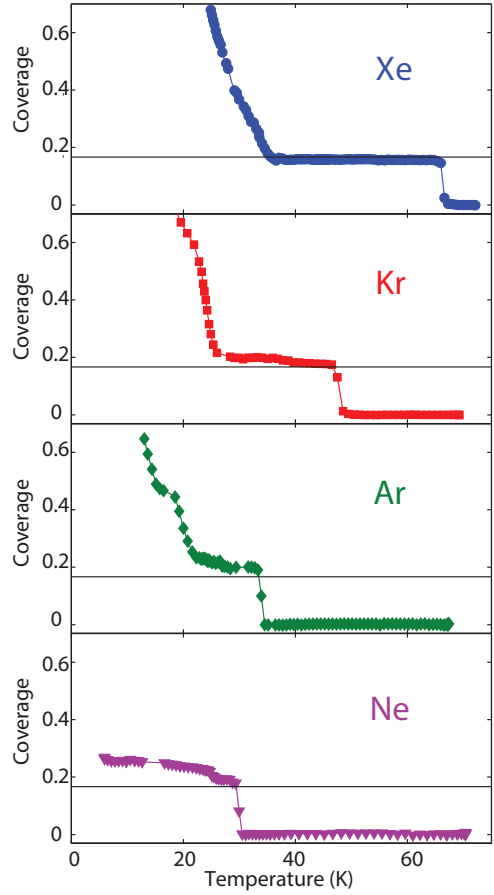}
      \caption{
   Coverage upon lowering temperature while dosing
   Xe, Kr, Ar, and Ne atoms. The pressure is $3\cdot 10^{-7}$ mbar for all measurements. 
   The $T$ ramping rate is $0.008\;\rm K/ s$ for the Xe measurement  
   and $0.016\;\rm K/ s$ for the other measurements. The black line corresponds to
   $\varphi =1/6$. }
      \label{fig:monolayers}
      \end{figure}

We grew different monolayers on the nanotube by dosing Xe, Kr, Ar,
and Ne (Fig. \ref{fig:monolayers}). The nanotube surface was
cleaned by current annealing before each growth. Upon decreasing
temperature, the coverage increases rapidly from 0 to a plateau
with $\varphi \simeq 1/6$, indicating the growth of the monolayer.
The characteristic temperature of the monolayer growth depends on
the atomic species; $T_c$ is higher when the atomic mass is larger
(Fig. \ref{fig:monolayers}). We attribute the origin of the
variation of $T_c$ to the polarizability of the atomic
species and the van der Waals interaction between the atom and the nanotube; 
the polarizability and the interaction both increasing with the atomic radius.
We also carried out experiments where we evaporated the monolayers
from the nanotube by continuously increasing the temperature of 
the cryostat from $4$ to $\sim$100 K. The
coverage jumped from $\simeq1/6$ to 0 at a temperature that is up
to $\sim$10 K higher than $T_c$ (Supplemental Material, Sec. V). 

We measured the time of the growth of monolayers
from $\varphi=0$ to $\varphi=1/6$ while keeping the temperature constant.
This time gets longer for lower pressure (Supplementary Material, sec. VI).

\begin{figure}[t!]
    \centering
    \includegraphics[scale=1]{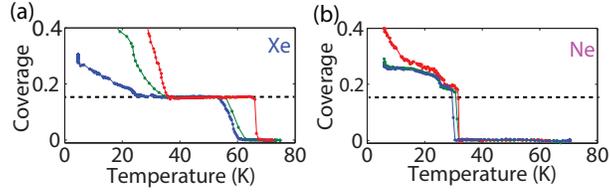}
    \caption{
    (a) Coverage upon lowering temperature while
   dosing Xe atoms. The pressure is   $7\cdot 10^{-8}$, $3\cdot 10^{-7}$, and $3\cdot 10^{-7}$ mbar, and the T
   ramping rate is $0.016$, $0.033$, and $0.008\;\rm K/s$ for the blue, green, and red
   lines, respectively. (b) Coverage upon lowering temperature while
   dosing Ne atoms. The pressure is $3\cdot 10^{-7}$ mbar for all three measurements, and the T
   ramping rate is $0.016\;\rm K/s$ for the blue and the green lines and $0.008\;\rm K/s$ for the red
   line.}
    \label{fig:Xe_Ne}
    \end{figure}

Xenon monolayers are particularly robust against temperature
changes. Figure \ref{fig:Xe_Ne}(a) shows coverage-temperature
measurements recorded at different pressures and different
temperature ramping rates. $T_{\rm c}$ varies from one measurement
to the next. However, the plateau in coverage at $1/6$ is clearly
reproducible. This shows that Xe monolayers are energetically
stable with the number of atoms being insensitive to temperature
over a large parameter space. In contrast, measurements with Ne
feature a plateau whose coverage depends significantly on $T$
(Fig. \ref{fig:Xe_Ne}(b)).

We now discuss the nature of the monolayers of Xe, Kr, Ar, and Ne.
For this, we carried out theoretical calculations to predict
whether the solid phases are commensurate or incommensurate in the
limit of zero temperature. In addition, we estimated the melting
temperature of the different solid phases. To this end, we
performed a series of Monte Carlo simulations relying on standard
interatomic potentials between noble gas atoms and the carbon
atoms of the nanotube (Supplementary material, Sec. IX). This
microscopic study was carried out for
nanotubes with diameters in the range 21--38 \AA, which covers the
typical diameters obtained with our chemical vapor deposition
recipe.

Our experimental findings indicate that Xe monolayers are
commensurate solids. Firstly, the coverage of the monolayer is
$1/6$. Secondly, the coverage remains at this value over a large
temperature range. This robustness suggests that Xe atoms are
strongly bound to the underlying carbon surface, as it is the case
for commensurate solids. Our experimental results are accounted
for by our theoretical calculations, which predict that the solid
is a registered $\sqrt{3} \times \sqrt{3}$ crystal at zero
temperature. Moreover, this solid phase is calculated to become
unstable at $\sim80$ K, which is consistent with the melting
temperature measured in Fig. S4.

Monolayers of Ar and Ne are less stable, since the measured
coverage depends significantly on temperature for $T \lesssim T_{\rm c}$. Our
calculations reveal that in the limit of zero temperature Ar and Ne 
monolayers are incommensurate solids with coverages 0.265 and 0.403,
respectively. The measured coverages at $T\lesssim T_{\rm c}$
 are much lower than these predicted values, suggesting that the monolayers
 observed experimentally are not in the solid phase. Moreover, 
our calculations show that incommensurate solids
melt at temperatures as low as 5~K when the coverage is set to the values we 
typically measure at $T\lesssim T_{\rm c}$. 
This further indicates that the monolayers of Ar 
and Ne observed experimentally at $25-35$ K  are in the liquid phase.

As for Kr, the measured temperature dependence of the coverage is
similar to that of Xe, supporting the scenario that Kr monolayers
are commensurate solid phases. This result would be in agreement
with experimental signatures of stability of a commensurate Kr
layer on graphite up to quite high temperatures, $T \sim
130$~K~\cite{Specht1987}. However, the coverage of Kr slightly
depends on temperature in the plateau region (Fig.~3), showing
that Kr monolayers are less pinned to the carbon surface than Xe
monolayers. Previous theoretical calculations of the Kr monolayer
on graphite show that corrugation effects are extremely important
to get the $\sqrt{3} \times \sqrt{3}$ crystal
stable~\cite{Vidali1984,Shrimpton1991}. Only by increasing in an
empirical way the anisotropic part of the pair interaction the
commensurate phase becomes stable. Our present simulations on Kr
adsorbed on nanotubes show the same trend. Therefore, more work is
needed to establish the phase of Kr monolayers on nanotubes.

To conclude, we studied the formation of noble gas atom monolayers
on individual nanotubes. We found that Xe atoms form robust
commensurate solids, whereas Ar and Ne atoms organize themselves
in fluids. These monolayers consist of $\sim 10^5$ atoms, which is
a tiny amount of material difficult to detect with most
experimental techniques used in surface science. The study of
these monolayers was here possible, because nanotube mechanical
resonators are extremely sensitive probes. The second important
aspect of our experiments is that the nanotube surface was
ultra-clean; this was achieved by thoroughly current annealing the
nanotube in ultra-high vacuum. These resonators made from
ultra-clean nanotubes are promising for various future adsorption
experiments, such as the measurement of new phase transitions
emerging in the one-dimensional limit with narrower nanotubes, the
investigation of quantum effects of He monolayers adsorbed on
nanotubes \cite{gordillo20124}, the study of the diffusion of
adsorbed atoms over the resonator surface which is a topic of increasing
interest \cite{yang2011surface}, and the interplay between the
strong mechanical nonlinearities of nanotubes
\cite{eichler2011nonlinear,eichler2012strong,eichler2013symmetry,Steele_PRB}
and the diffusion of atoms
\cite{Atalaya:2011jh,Dykman_PRB,barnard2012fluctuation}.

We thank A. Isacsson and J. Moser for discussions. We acknowledge
support from the European Union through the Graphene Flagship (604391), the
ERC-carbonNEMS project, and a Marie Curie grant (271938), the
Spanish state (MAT2012-31338), and the Catalan
government (AGAUR, SGR). C. G. and J. B.  acknowledge partial
financial support from the Junta de de Andaluc\'{\i}a Group
PAI-205, Grant No. FQM-5987, MICINN (Spain) Grants No.
FIS2010-18356 and FIS2011-25275, and Generalitat de Catalunya
Grant 2009SGR-1003.

\bibliography{references}

\begin{thebibliography}{28}
\expandafter\ifx\csname natexlab\endcsname\relax\def\natexlab#1{#1}\fi
\expandafter\ifx\csname bibnamefont\endcsname\relax
  \def\bibnamefont#1{#1}\fi
\expandafter\ifx\csname bibfnamefont\endcsname\relax
  \def\bibfnamefont#1{#1}\fi
\expandafter\ifx\csname citenamefont\endcsname\relax
  \def\citenamefont#1{#1}\fi
\expandafter\ifx\csname url\endcsname\relax
  \def\url#1{\texttt{#1}}\fi
\expandafter\ifx\csname urlprefix\endcsname\relax\def\urlprefix{URL }\fi
\providecommand{\bibinfo}[2]{#2}
\providecommand{\eprint}[2][]{\url{#2}}

\bibitem[{\citenamefont{Dillon et~al.}(1997)\citenamefont{Dillon, Jones,
  Bekkedahl, Kiang, Bethune, and Heben}}]{dillon1997storage}
\bibinfo{author}{\bibfnamefont{A.}~\bibnamefont{Dillon}},
  \bibinfo{author}{\bibfnamefont{K.}~\bibnamefont{Jones}},
  \bibinfo{author}{\bibfnamefont{T.}~\bibnamefont{Bekkedahl}},
  \bibinfo{author}{\bibfnamefont{C.}~\bibnamefont{Kiang}},
  \bibinfo{author}{\bibfnamefont{D.}~\bibnamefont{Bethune}}, \bibnamefont{and}
  \bibinfo{author}{\bibfnamefont{M.}~\bibnamefont{Heben}},
  \bibinfo{journal}{Nature} \textbf{\bibinfo{volume}{386}},
  \bibinfo{pages}{377} (\bibinfo{year}{1997}).

\bibitem[{\citenamefont{Teizer et~al.}(1999)\citenamefont{Teizer, Hallock,
  Dujardin, and Ebbesen}}]{teizer19994}
\bibinfo{author}{\bibfnamefont{W.}~\bibnamefont{Teizer}},
  \bibinfo{author}{\bibfnamefont{R.}~\bibnamefont{Hallock}},
  \bibinfo{author}{\bibfnamefont{E.}~\bibnamefont{Dujardin}}, \bibnamefont{and}
  \bibinfo{author}{\bibfnamefont{T.}~\bibnamefont{Ebbesen}},
  \bibinfo{journal}{Physical Review Letters} \textbf{\bibinfo{volume}{82}},
  \bibinfo{pages}{5305} (\bibinfo{year}{1999}).

\bibitem[{\citenamefont{Ulbricht et~al.}(2002)\citenamefont{Ulbricht, Kriebel,
  Moos, and Hertel}}]{ulbricht2002desorption}
\bibinfo{author}{\bibfnamefont{H.}~\bibnamefont{Ulbricht}},
  \bibinfo{author}{\bibfnamefont{J.}~\bibnamefont{Kriebel}},
  \bibinfo{author}{\bibfnamefont{G.}~\bibnamefont{Moos}}, \bibnamefont{and}
  \bibinfo{author}{\bibfnamefont{T.}~\bibnamefont{Hertel}},
  \bibinfo{journal}{Chemical Physics Letters} \textbf{\bibinfo{volume}{363}},
  \bibinfo{pages}{252} (\bibinfo{year}{2002}).

\bibitem[{\citenamefont{Shi and Johnson}(2003)}]{shi2003gas}
\bibinfo{author}{\bibfnamefont{W.}~\bibnamefont{Shi}} \bibnamefont{and}
  \bibinfo{author}{\bibfnamefont{J.~K.} \bibnamefont{Johnson}},
  \bibinfo{journal}{Physical Review Letters} \textbf{\bibinfo{volume}{91}},
  \bibinfo{pages}{015504} (\bibinfo{year}{2003}).

\bibitem[{\citenamefont{Lasjaunias et~al.}(2003)\citenamefont{Lasjaunias,
  Biljakovi{\'c}, Sauvajol, and Monceau}}]{lasjaunias2003evidence}
\bibinfo{author}{\bibfnamefont{J.-C.} \bibnamefont{Lasjaunias}},
  \bibinfo{author}{\bibfnamefont{K.}~\bibnamefont{Biljakovi{\'c}}},
  \bibinfo{author}{\bibfnamefont{J.-L.} \bibnamefont{Sauvajol}},
  \bibnamefont{and} \bibinfo{author}{\bibfnamefont{P.}~\bibnamefont{Monceau}},
  \bibinfo{journal}{Physical Review Letters} \textbf{\bibinfo{volume}{91}},
  \bibinfo{pages}{025901} (\bibinfo{year}{2003}).

\bibitem[{\citenamefont{Ulbricht et~al.}(2006)\citenamefont{Ulbricht, Zacharia,
  Cindir, and Hertel}}]{ulbricht2006thermal}
\bibinfo{author}{\bibfnamefont{H.}~\bibnamefont{Ulbricht}},
  \bibinfo{author}{\bibfnamefont{R.}~\bibnamefont{Zacharia}},
  \bibinfo{author}{\bibfnamefont{N.}~\bibnamefont{Cindir}}, \bibnamefont{and}
  \bibinfo{author}{\bibfnamefont{T.}~\bibnamefont{Hertel}},
  \bibinfo{journal}{Carbon} \textbf{\bibinfo{volume}{44}},
  \bibinfo{pages}{2931} (\bibinfo{year}{2006}).

\bibitem[{\citenamefont{Wang et~al.}(2010)\citenamefont{Wang, Wei, Morse, Dash,
  Vilches, and Cobden}}]{Wang:2010hi}
\bibinfo{author}{\bibfnamefont{Z.}~\bibnamefont{Wang}},
  \bibinfo{author}{\bibfnamefont{J.}~\bibnamefont{Wei}},
  \bibinfo{author}{\bibfnamefont{P.}~\bibnamefont{Morse}},
  \bibinfo{author}{\bibfnamefont{J.~G.} \bibnamefont{Dash}},
  \bibinfo{author}{\bibfnamefont{O.~E.} \bibnamefont{Vilches}},
  \bibnamefont{and} \bibinfo{author}{\bibfnamefont{D.~H.}
  \bibnamefont{Cobden}}, \bibinfo{journal}{Science}
  \textbf{\bibinfo{volume}{327}}, \bibinfo{pages}{552} (\bibinfo{year}{2010}).

\bibitem[{\citenamefont{Lee et~al.}(2012)\citenamefont{Lee, Vilches, Wang,
  Fredrickson, Morse, Roy, Dzyubenko, and Cobden}}]{Lee:2012ic}
\bibinfo{author}{\bibfnamefont{H.-C.} \bibnamefont{Lee}},
  \bibinfo{author}{\bibfnamefont{O.~E.} \bibnamefont{Vilches}},
  \bibinfo{author}{\bibfnamefont{Z.}~\bibnamefont{Wang}},
  \bibinfo{author}{\bibfnamefont{E.}~\bibnamefont{Fredrickson}},
  \bibinfo{author}{\bibfnamefont{P.}~\bibnamefont{Morse}},
  \bibinfo{author}{\bibfnamefont{R.}~\bibnamefont{Roy}},
  \bibinfo{author}{\bibfnamefont{B.}~\bibnamefont{Dzyubenko}},
  \bibnamefont{and} \bibinfo{author}{\bibfnamefont{D.~H.}
  \bibnamefont{Cobden}}, \bibinfo{journal}{Journal of Low Temperature Physics}
  \textbf{\bibinfo{volume}{169}}, \bibinfo{pages}{338} (\bibinfo{year}{2012}).

\bibitem[{\citenamefont{Chaste et~al.}(2012)\citenamefont{Chaste, Eichler,
  Moser, Ceballos, Rurali, and Bachtold}}]{Chaste_NatNano12}
\bibinfo{author}{\bibfnamefont{J.}~\bibnamefont{Chaste}},
  \bibinfo{author}{\bibfnamefont{A.}~\bibnamefont{Eichler}},
  \bibinfo{author}{\bibfnamefont{J.}~\bibnamefont{Moser}},
  \bibinfo{author}{\bibfnamefont{G.}~\bibnamefont{Ceballos}},
  \bibinfo{author}{\bibfnamefont{R.}~\bibnamefont{Rurali}}, \bibnamefont{and}
  \bibinfo{author}{\bibfnamefont{A.}~\bibnamefont{Bachtold}},
  \bibinfo{journal}{Nature Nanotechnology} \textbf{\bibinfo{volume}{7}},
  \bibinfo{pages}{301} (\bibinfo{year}{2012}).

\bibitem[{\citenamefont{Chiu et~al.}(2008)\citenamefont{Chiu, Hung, Postma, and
  Bockrath}}]{Chiu:2008eq}
\bibinfo{author}{\bibfnamefont{H.-Y.} \bibnamefont{Chiu}},
  \bibinfo{author}{\bibfnamefont{P.}~\bibnamefont{Hung}},
  \bibinfo{author}{\bibfnamefont{H.~W.~C.} \bibnamefont{Postma}},
  \bibnamefont{and} \bibinfo{author}{\bibfnamefont{M.}~\bibnamefont{Bockrath}},
  \bibinfo{journal}{Nano Letters} \textbf{\bibinfo{volume}{8}},
  \bibinfo{pages}{4342} (\bibinfo{year}{2008}).

\bibitem[{\citenamefont{Lassagne et~al.}(2008)\citenamefont{Lassagne,
  Garcia-Sanchez, Aguasca, and Bachtold}}]{lassagne2008ultrasensitive}
\bibinfo{author}{\bibfnamefont{B.}~\bibnamefont{Lassagne}},
  \bibinfo{author}{\bibfnamefont{D.}~\bibnamefont{Garcia-Sanchez}},
  \bibinfo{author}{\bibfnamefont{A.}~\bibnamefont{Aguasca}}, \bibnamefont{and}
  \bibinfo{author}{\bibfnamefont{A.}~\bibnamefont{Bachtold}},
  \bibinfo{journal}{Nano Letters} \textbf{\bibinfo{volume}{8}},
  \bibinfo{pages}{3735} (\bibinfo{year}{2008}).

\bibitem[{\citenamefont{Jensen et~al.}(2008)\citenamefont{Jensen, Kim, and
  Zettl}}]{jensen2008atomic}
\bibinfo{author}{\bibfnamefont{K.}~\bibnamefont{Jensen}},
  \bibinfo{author}{\bibfnamefont{K.}~\bibnamefont{Kim}}, \bibnamefont{and}
  \bibinfo{author}{\bibfnamefont{A.}~\bibnamefont{Zettl}},
  \bibinfo{journal}{Nature Nanotechnology} \textbf{\bibinfo{volume}{3}},
  \bibinfo{pages}{533} (\bibinfo{year}{2008}).

\bibitem[{\citenamefont{Bruch et~al.}(1997)\citenamefont{Bruch, Cole, and
  Zaremba}}]{bruch1997physical}
\bibinfo{author}{\bibfnamefont{L.~W.} \bibnamefont{Bruch}},
  \bibinfo{author}{\bibfnamefont{M.~W.} \bibnamefont{Cole}}, \bibnamefont{and}
  \bibinfo{author}{\bibfnamefont{E.}~\bibnamefont{Zaremba}},
  \emph{\bibinfo{title}{Physical adsorption: forces and phenomena}}
  (\bibinfo{publisher}{Clarendon Press Oxford}, \bibinfo{year}{1997}).

\bibitem[{\citenamefont{Eichler et~al.}(2011)\citenamefont{Eichler, Moser,
  Chaste, Zdrojek, Wilson-Rae, and Bachtold}}]{eichler2011nonlinear}
\bibinfo{author}{\bibfnamefont{A.}~\bibnamefont{Eichler}},
  \bibinfo{author}{\bibfnamefont{J.}~\bibnamefont{Moser}},
  \bibinfo{author}{\bibfnamefont{J.}~\bibnamefont{Chaste}},
  \bibinfo{author}{\bibfnamefont{M.}~\bibnamefont{Zdrojek}},
  \bibinfo{author}{\bibfnamefont{I.}~\bibnamefont{Wilson-Rae}},
  \bibnamefont{and} \bibinfo{author}{\bibfnamefont{A.}~\bibnamefont{Bachtold}},
  \bibinfo{journal}{Nature Nanotechnology} \textbf{\bibinfo{volume}{6}},
  \bibinfo{pages}{339} (\bibinfo{year}{2011}).

\bibitem[{\citenamefont{Huttel et~al.}(2009)\citenamefont{Huttel, Steele,
  Witkamp, Poot, Kouwenhoven, and van~der Zant}}]{huttel2009carbon}
\bibinfo{author}{\bibfnamefont{A.~K.} \bibnamefont{Huttel}},
  \bibinfo{author}{\bibfnamefont{G.~A.} \bibnamefont{Steele}},
  \bibinfo{author}{\bibfnamefont{B.}~\bibnamefont{Witkamp}},
  \bibinfo{author}{\bibfnamefont{M.}~\bibnamefont{Poot}},
  \bibinfo{author}{\bibfnamefont{L.~P.} \bibnamefont{Kouwenhoven}},
  \bibnamefont{and} \bibinfo{author}{\bibfnamefont{H.~S.} \bibnamefont{van~der
  Zant}}, \bibinfo{journal}{Nano Letters} \textbf{\bibinfo{volume}{9}},
  \bibinfo{pages}{2547} (\bibinfo{year}{2009}).

\bibitem[{\citenamefont{Gouttenoire et~al.}(2010)\citenamefont{Gouttenoire,
  Barois, Perisanu, Leclercq, Purcell, Vincent, and Ayari}}]{Barois_Small10}
\bibinfo{author}{\bibfnamefont{V.}~\bibnamefont{Gouttenoire}},
  \bibinfo{author}{\bibfnamefont{T.}~\bibnamefont{Barois}},
  \bibinfo{author}{\bibfnamefont{S.}~\bibnamefont{Perisanu}},
  \bibinfo{author}{\bibfnamefont{J.-L.} \bibnamefont{Leclercq}},
  \bibinfo{author}{\bibfnamefont{S.~T.} \bibnamefont{Purcell}},
  \bibinfo{author}{\bibfnamefont{P.}~\bibnamefont{Vincent}}, \bibnamefont{and}
  \bibinfo{author}{\bibfnamefont{A.}~\bibnamefont{Ayari}},
  \bibinfo{journal}{Small} \textbf{\bibinfo{volume}{6}}, \bibinfo{pages}{1060}
  (\bibinfo{year}{2010}).

\bibitem[{\citenamefont{Chaste et~al.}(2011)\citenamefont{Chaste, Sledzinska,
  Zdrojek, Moser, and Bachtold}}]{Chaste:2011bj}
\bibinfo{author}{\bibfnamefont{J.}~\bibnamefont{Chaste}},
  \bibinfo{author}{\bibfnamefont{M.}~\bibnamefont{Sledzinska}},
  \bibinfo{author}{\bibfnamefont{M.}~\bibnamefont{Zdrojek}},
  \bibinfo{author}{\bibfnamefont{J.}~\bibnamefont{Moser}}, \bibnamefont{and}
  \bibinfo{author}{\bibfnamefont{A.}~\bibnamefont{Bachtold}},
  \bibinfo{journal}{Applied Physics Letters} \textbf{\bibinfo{volume}{99}},
  \bibinfo{pages}{213502} (\bibinfo{year}{2011}).

\bibitem[{\citenamefont{Specht et~al.}(1987)\citenamefont{Specht, Mak, Peters,
  Sutton, Birgeneau, D'Amico, Moncton, Nagler, and Horn}}]{Specht1987}
\bibinfo{author}{\bibfnamefont{E.~D.} \bibnamefont{Specht}},
  \bibinfo{author}{\bibfnamefont{A.}~\bibnamefont{Mak}},
  \bibinfo{author}{\bibfnamefont{C.}~\bibnamefont{Peters}},
  \bibinfo{author}{\bibfnamefont{M.}~\bibnamefont{Sutton}},
  \bibinfo{author}{\bibfnamefont{R.~J.} \bibnamefont{Birgeneau}},
  \bibinfo{author}{\bibfnamefont{D.~L.} \bibnamefont{D'Amico}},
  \bibinfo{author}{\bibfnamefont{D.~E.} \bibnamefont{Moncton}},
  \bibinfo{author}{\bibfnamefont{S.~E.} \bibnamefont{Nagler}},
  \bibnamefont{and} \bibinfo{author}{\bibfnamefont{P.~M.} \bibnamefont{Horn}},
  \bibinfo{journal}{Z. Phys. B} \textbf{\bibinfo{volume}{69}},
  \bibinfo{pages}{347} (\bibinfo{year}{1987}).

\bibitem[{\citenamefont{Vidali and Cole}(1984)}]{Vidali1984}
\bibinfo{author}{\bibfnamefont{G.}~\bibnamefont{Vidali}} \bibnamefont{and}
  \bibinfo{author}{\bibfnamefont{M.~W.} \bibnamefont{Cole}},
  \bibinfo{journal}{Phys. Rev. B} \textbf{\bibinfo{volume}{29}},
  \bibinfo{pages}{6736} (\bibinfo{year}{1984}).

\bibitem[{\citenamefont{Shrimpton and W.~A.~Steele}(1991)}]{Shrimpton1991}
\bibinfo{author}{\bibfnamefont{N.~D.} \bibnamefont{Shrimpton}}
  \bibnamefont{and} \bibinfo{author}{\bibfnamefont{W.~A.}
  \bibnamefont{W.~A.~Steele}}, \bibinfo{journal}{Phys. Rev. B}
  \textbf{\bibinfo{volume}{44}}, \bibinfo{pages}{3297} (\bibinfo{year}{1991}).

\bibitem[{\citenamefont{Gordillo and Boronat}(2012)}]{gordillo20124}
\bibinfo{author}{\bibfnamefont{M.}~\bibnamefont{Gordillo}} \bibnamefont{and}
  \bibinfo{author}{\bibfnamefont{J.}~\bibnamefont{Boronat}},
  \bibinfo{journal}{Physical Review B} \textbf{\bibinfo{volume}{86}},
  \bibinfo{pages}{165409} (\bibinfo{year}{2012}).

\bibitem[{\citenamefont{Yang et~al.}(2011)\citenamefont{Yang, Callegari, Feng,
  and Roukes}}]{yang2011surface}
\bibinfo{author}{\bibfnamefont{Y.}~\bibnamefont{Yang}},
  \bibinfo{author}{\bibfnamefont{C.}~\bibnamefont{Callegari}},
  \bibinfo{author}{\bibfnamefont{X.}~\bibnamefont{Feng}}, \bibnamefont{and}
  \bibinfo{author}{\bibfnamefont{M.}~\bibnamefont{Roukes}},
  \bibinfo{journal}{Nano Letters} \textbf{\bibinfo{volume}{11}},
  \bibinfo{pages}{1753} (\bibinfo{year}{2011}).

\bibitem[{\citenamefont{Eichler et~al.}(2012)\citenamefont{Eichler, del
  {\'A}lamo~Ruiz, Plaza, and Bachtold}}]{eichler2012strong}
\bibinfo{author}{\bibfnamefont{A.}~\bibnamefont{Eichler}},
  \bibinfo{author}{\bibfnamefont{M.}~\bibnamefont{del {\'A}lamo~Ruiz}},
  \bibinfo{author}{\bibfnamefont{J.}~\bibnamefont{Plaza}}, \bibnamefont{and}
  \bibinfo{author}{\bibfnamefont{A.}~\bibnamefont{Bachtold}},
  \bibinfo{journal}{Physical Review Letters} \textbf{\bibinfo{volume}{109}},
  \bibinfo{pages}{025503} (\bibinfo{year}{2012}).

\bibitem[{\citenamefont{Eichler et~al.}(2013)\citenamefont{Eichler, Moser,
  Dykman, and Bachtold}}]{eichler2013symmetry}
\bibinfo{author}{\bibfnamefont{A.}~\bibnamefont{Eichler}},
  \bibinfo{author}{\bibfnamefont{J.}~\bibnamefont{Moser}},
  \bibinfo{author}{\bibfnamefont{M.}~\bibnamefont{Dykman}}, \bibnamefont{and}
  \bibinfo{author}{\bibfnamefont{A.}~\bibnamefont{Bachtold}},
  \bibinfo{journal}{Nature Communications} \textbf{\bibinfo{volume}{4}},
  \bibinfo{pages}{2843} (\bibinfo{year}{2013}).

\bibitem[{\citenamefont{Meerwaldt et~al.}(2012)\citenamefont{Meerwaldt,
  Labadze, Schneider, Taspinar, Blanter, van~der Zant, and
  Steele}}]{Steele_PRB}
\bibinfo{author}{\bibfnamefont{H.~B.} \bibnamefont{Meerwaldt}},
  \bibinfo{author}{\bibfnamefont{G.}~\bibnamefont{Labadze}},
  \bibinfo{author}{\bibfnamefont{B.~H.} \bibnamefont{Schneider}},
  \bibinfo{author}{\bibfnamefont{A.}~\bibnamefont{Taspinar}},
  \bibinfo{author}{\bibfnamefont{Y.~M.} \bibnamefont{Blanter}},
  \bibinfo{author}{\bibfnamefont{H.~S.~J.} \bibnamefont{van~der Zant}},
  \bibnamefont{and} \bibinfo{author}{\bibfnamefont{G.~A.}
  \bibnamefont{Steele}}, \bibinfo{journal}{Phys. Rev. B}
  \textbf{\bibinfo{volume}{86}}, \bibinfo{pages}{115454}
  (\bibinfo{year}{2012}).

\bibitem[{\citenamefont{Atalaya
  et~al.}(2011{\natexlab{a}})\citenamefont{Atalaya, Isacsson, and
  Dykman}}]{Atalaya:2011jh}
\bibinfo{author}{\bibfnamefont{J.}~\bibnamefont{Atalaya}},
  \bibinfo{author}{\bibfnamefont{A.}~\bibnamefont{Isacsson}}, \bibnamefont{and}
  \bibinfo{author}{\bibfnamefont{M.}~\bibnamefont{Dykman}},
  \bibinfo{journal}{Physical Review Letters} \textbf{\bibinfo{volume}{106}},
  \bibinfo{pages}{227202} (\bibinfo{year}{2011}{\natexlab{a}}).

\bibitem[{\citenamefont{Atalaya
  et~al.}(2011{\natexlab{b}})\citenamefont{Atalaya, Isacsson, and
  Dykman}}]{Dykman_PRB}
\bibinfo{author}{\bibfnamefont{J.}~\bibnamefont{Atalaya}},
  \bibinfo{author}{\bibfnamefont{A.}~\bibnamefont{Isacsson}}, \bibnamefont{and}
  \bibinfo{author}{\bibfnamefont{M.~I.} \bibnamefont{Dykman}},
  \bibinfo{journal}{Phys. Rev. B} \textbf{\bibinfo{volume}{83}},
  \bibinfo{pages}{045419} (\bibinfo{year}{2011}{\natexlab{b}}).

\bibitem[{\citenamefont{Barnard et~al.}(2012)\citenamefont{Barnard, Sazonova,
  van~der Zande, and McEuen}}]{barnard2012fluctuation}
\bibinfo{author}{\bibfnamefont{A.~W.} \bibnamefont{Barnard}},
  \bibinfo{author}{\bibfnamefont{V.}~\bibnamefont{Sazonova}},
  \bibinfo{author}{\bibfnamefont{A.~M.} \bibnamefont{van~der Zande}},
  \bibnamefont{and} \bibinfo{author}{\bibfnamefont{P.~L.}
  \bibnamefont{McEuen}}, \bibinfo{journal}{Proceedings of the National Academy
  of Sciences} \textbf{\bibinfo{volume}{109}}, \bibinfo{pages}{19093}
  (\bibinfo{year}{2012}).

\end{thebibliography}
\newpage

\end{document}